\documentclass{elsart}
\usepackage[utf8]{inputenc} 
\usepackage{graphicx}
\usepackage{subcaption}
\usepackage{amssymb}
\usepackage{mathtools} 
\usepackage{amsmath}
\usepackage{color}
\usepackage[dvipsnames]{xcolor}
\usepackage{bm}
\usepackage{cite}
\usepackage{float}

\usepackage{color}
\usepackage{ulem}
\usepackage{multirow}

\begin{document}

\begin{frontmatter}

\title{Study of the threshold anomaly in the elastic scattering of d + $^{197}$Au}

\author[1]{T. Giudice,}
\author[1]{D. Abriola,}
\author[1,2]{A. Arazi,}
\author[1]{E. de Barbará,}
\author[1,2]{M.A. Cardona,}
\author[2]{J. G\'omez,}
\author[1,2]{D. Hojman,}
\author[3,4]{R.M. Id Betan,}
\author[1]{M.S. Kohen,}
\author[4]{N. Llaneza,}
\author[1]{G.V. Martí,}
\author[1]{B. Paes,}
\author[1,2]{D. Schneider,}
\author[5]{H. Soler,}
\author[5]{J. Lubian}

\address[1]{Laboratorio TANDAR, Comisión Nacional de Energía Atómica, BKNA1650 San Martín, Argentina}
\address[2]{Consejo Nacional de Investigaciones Científicas y Técnicas, C1425FQB Buenos Aires, Argentina}
\address[3]{Instituto de Física de Rosario (CONICET-UNR), Bv. 27 de Febrero 210 bis, S2000EZP Rosario, Argentina}
\address[4]{Facultad de Ciencias Exactas, Ingeniería y Agrimensura (UNR), Av. Pellegrini 250, S2000BTP Rosario, Argentina}
\address[5]{Instituto de F\'isica, Universidade Federal Fluminense, Niteroi, Brasil}
                             
\date{\today}

\begin{abstract}
Measurements of the elastic scattering angular distribution for the d + $^{197}$Au system were carried out covering deuteron incident energies in the range  from 5 to 16~MeV, i.e. approximately 50\% below and above the Coulomb barrier. A  critical interaction distance of $d_I= 2.49$~fm was determined from these  distributions, which is  comparable to that of the radioactive halo nucleus $^6$He. The experimental angular distributions were systematically analyzed using two alternative models: the semi-microscopic São Paulo and the effective Woods-Saxon optical potentials, for which the best-fitting parameters were determined. These  potentials, integrated in the vicinity of the sensitivity radius, were calculated for each energy. For both models, the energy dependence of these integrals  presented the breakup threshold anomaly around the coulomb barrier, a typical signature of weakly bound nuclei.
\end{abstract}

  
\end{frontmatter}

\section{Introduction}\label{sec:intro}
Elastic scattering is a source of relevant information about reaction mechanisms. The angular distribution of the differential cross sections of elastically scattered particles is influenced by processes that absorb incoming flux from the elastic channel as inelastic, transfer, breakup,  fusion, or part of the fission channels. These processes can be phenomenologically considered by an Optical Model (OM) with an imaginary nuclear part representing those absorptive process. The real and imaginary components are, however, not independent. They are connected by the dispersion relation as a consequence of the causality principle. In reactions involving strongly bound nuclei, the real part of the optical potential exhibits a bell-shaped peak at energies around the Coulomb barrier, whereas the imaginary part decreases for energies below this barrier \cite{mns86}. Named threshold anomaly (TA) \cite{s91}, this phenomenon was interpreted as a signature of the closing of the non-elastic channels below the Coulomb barrier.  

On the contrary, weakly bound nuclei, which have low breakup energies, may show an increase of the imaginary component of the potential around the barrier, whereas the real part drops \cite{sar00,pal03,nfa07}. This effect was associated with the non-fully closing of the non-elastic channels, particularly the breakup and transfer channels. The impact of the breakup can be described by the action of a repulsive polarization potential which,  added to the optical one, inhibits the closing of the absorptive processes. Therefore, the name breakup threshold anomaly (BTA) was coined for this effect \cite{hgl07,cgd06}. Although it is far from conclusive, depending on the quality of the experimental data and the model being used for the nuclear potential, weakly bound projectiles tend to exhibit the BTA \cite{nfa07,ffm12,GAA18,GAA20}, while strongly bound projectiles present the usual TA \cite{ffa10}.

The BTA analysis has been reported by studying the energy dependence of the optical potential that accounts for the reactions of systems involving stable weakly bound projectiles like $^6$Li~\cite{FNA07,PAD04,BSR08,SLC07,ZFP09,ffm12,KJR08,DMP11,MGL99,GPF05,FNA10,KBC94,SKR11,DMB14}, $^7$Li~\cite{FAF06,DMB14,PAD04,SLC07,ffm12,DMN11,MGL99,LPG01,FNA10,KBC94,DMB14}, and $^9$Be~\cite{OCL11,GAM04,PSS14a,GLP09,sar00,WFC04}, as well as radioactive like $^6$He~\cite{GLP07,SEA08}, $^{7}$Be and $^8$B~\cite{GAM10}. In all these studies, it was found that the imaginary part of the optical potential does not drop to zero at energies centered on the Coulomb barrier, as expected when the usual TA is observed. This effect was attributed to the breakup and transfer channels that remained open at energies below the barrier. The study of the energy dependence of the optical potential of reactions involving deuteron as a projectile has been reported for the d+$^{208}$Pb system in Ref. \cite{MG99}. The authors found the usual threshold anomaly for the nuclear part of the polarization potential. The Coulomb polarization potential was considered dominated by dipole excitations, and it was determined following the procedure developed in Ref. \cite{agn95}. In the present work, we study the energy dependence of the full polarization potential by fitting elastic scattering angular distributions for the d+$^{197}$Au at near-barrier energies. The deuteron is the simplest nuclear laboratory to study nuclear reactions, so it would be interesting to investigate if the BTA is present for such a simple system, where the concurrence between the nuclear and Coulomb interactions should still be present.

This article presents the experimental data for the elastic scattering cross section of the $^2$H + $^{197}$Au system at ten incident energies, covering a wide range above and below the Coulomb barrier. Systematic analysis to study the threshold behavior of the deuteron is presented. The incident deuteron energies were chosen around the Coulomb barrier, which is estimated to be $V_{B}=10.5$ MeV at the center-of-mass system. Energy-dependent optimized Woods-Saxon (WSP) and S\~ao Paulo (SPP) optical potentials were used in the present calculations to study the TA. To our knowledge, this is the first study of the TA for the deuteron. 

This paper is organized as follows. In Sec. \ref{sec:experimental}, the experimental setup is described; in Sec. \ref{sec:elastic}, the critical distance and the optical potentials are introduced. Sec. \ref{sec.sr} shows the sensitivity radii calculations to be used to analyze the threshold behavior of the deuteron in Sec. \ref{sec.ta}. Finally, the summary and conclusions are overviewed  in Sec. \ref{sec:discussion}.

\section{Experimental setup and measured cross sections} \label{sec:experimental}
The $^2$H beams were extracted from a TiD$_2$ material in a cesium  sputtering ion source. These ions were  accelerated by the  20UD tandem accelerator TANDAR \cite{tandar} at laboratory energies of 5, 6, 8, 9, 9.5, 10, 11, 12, 12.5, and 16 MeV. The relative energy spread of the beam  was in the order of $10^{-4}$. The beam current was limited to few nanoampers to maintain the counting rate of the detection system below 500 Hz, preventing significant corrections due to dead-time effects. Self-supporting $^{197}$Au foils, with thicknesses between 350 and 1700~$\mu$g/cm$^2$, were mounted as targets in the center of a 70-cm-diameter scattering chamber. The target's normal was set at  40$^\circ$ ($-40^\circ$) relative to the beam direction, for the measurement at forward (backward) angles, respectively. These orientations prevent the blocking of the reaction ejectiles by the target or its holder.

\begin{figure}[h!t]
     \centering
         \includegraphics[scale=0.16]{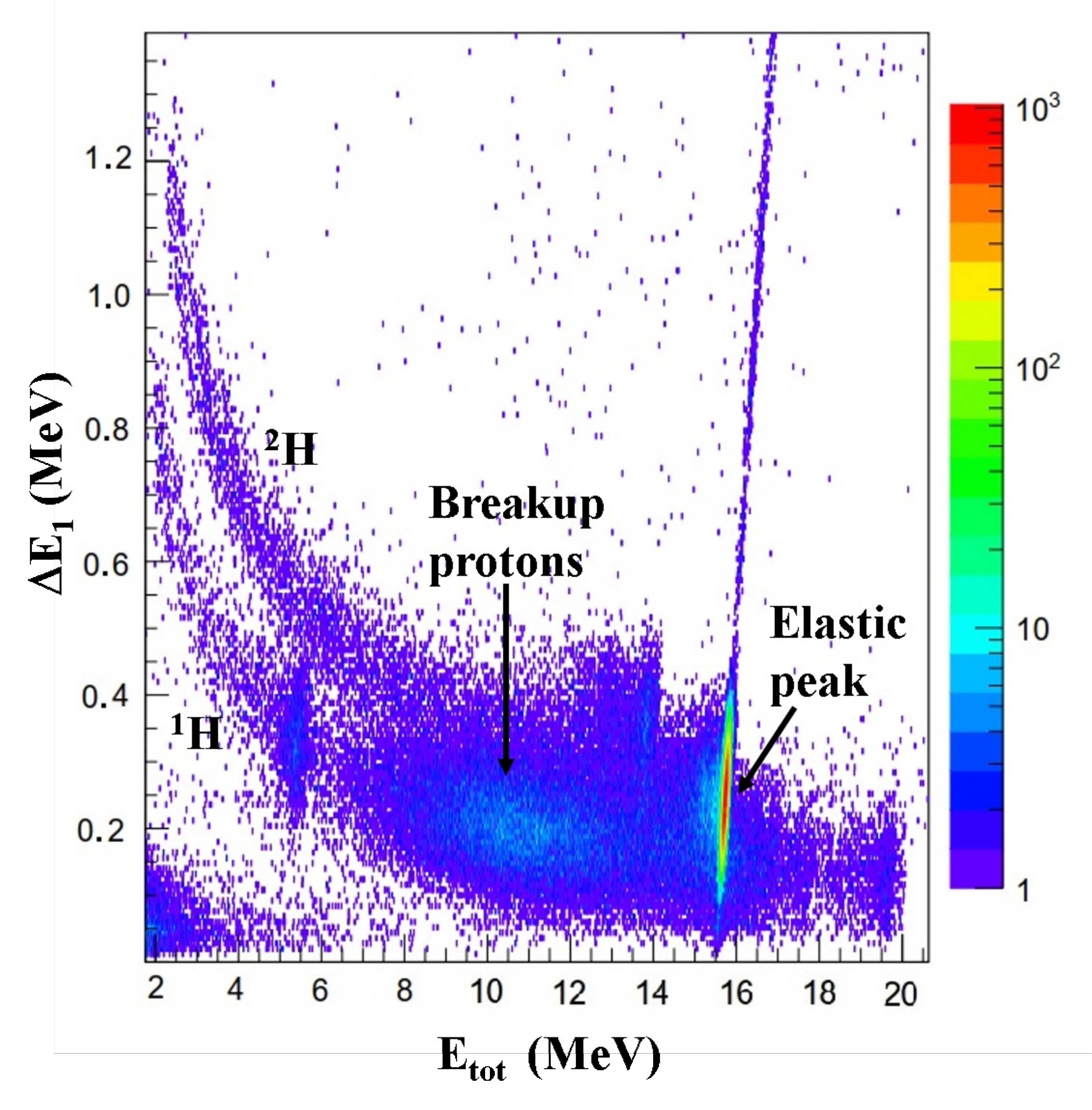}\hspace{3mm}
         \includegraphics[scale=0.16]{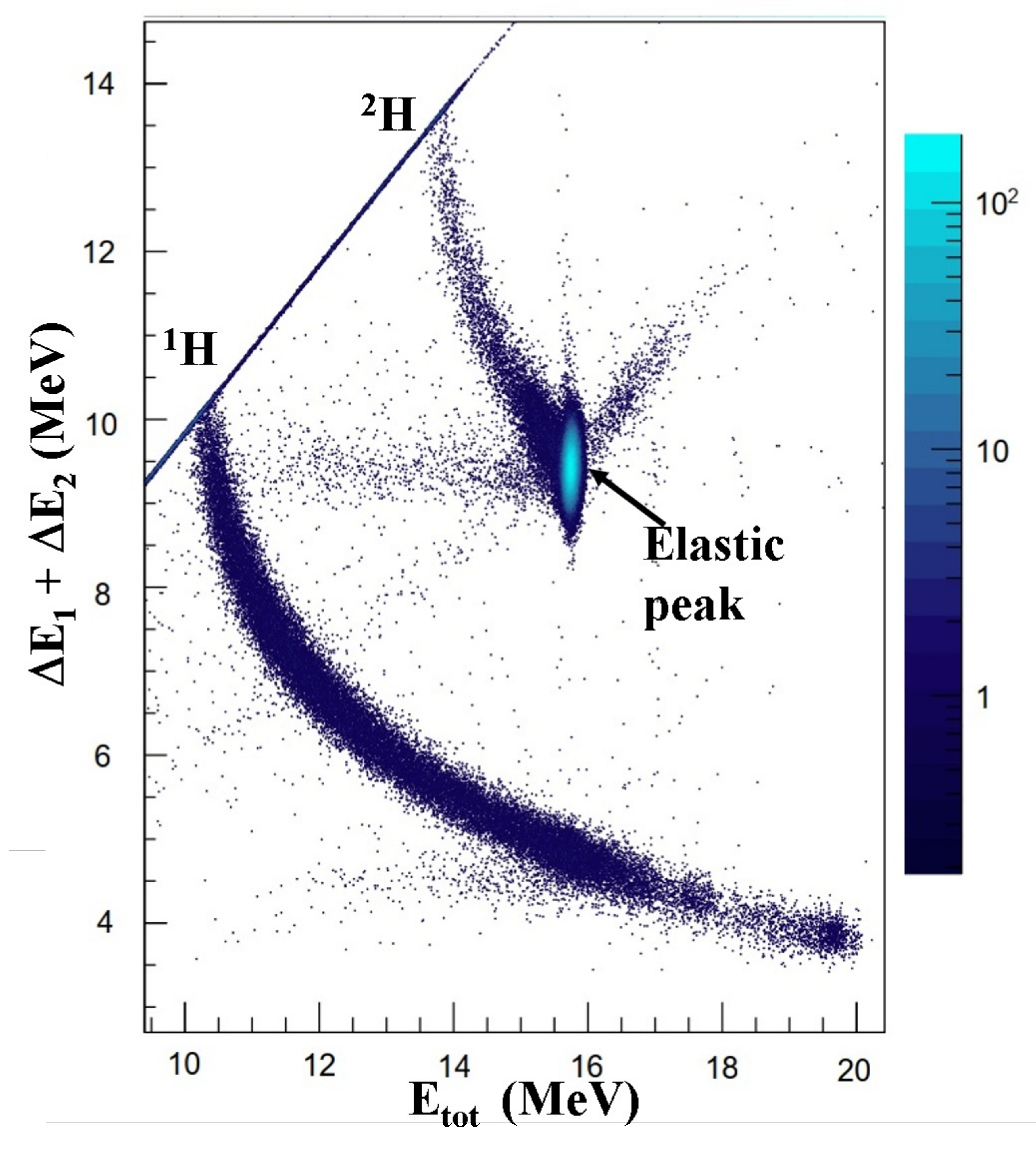}
    \caption{Two-dimensional spectra recorded at $\theta_{\rm lab} = 60^\circ$ for $E_{\rm lab} = 16$~MeV. Spectra of $\Delta E_1$ vs. $E_{\rm{tot}}$ (Left) and $\Delta E_1 + \Delta E_2$ vs. $E_{\rm{tot}}$ (Right).}
    \label{spectrum-2D}
\end{figure}

For particle detection and identification, two  telescopes were used. Each one consisted of four surface-barrier silicon detectors (with transmission mounting), named $\Delta E_1$,\dots,$\Delta E_4$, with increasing thicknesses: 22, 150, 700, and $1500~\mu\rm{m}$ for one telescope and  25, 700, 1000, and 1500 $\mu$m for the other. The total thickness ensure the stopping of protons with up to 20 MeV, such as those produced by the neutron stripping of 16~MeV deuterons. These telescopes were mounted on a rotatable disk at the base of the chamber, separated 20$^\circ$ from each other; and placed at a distance of  185 mm from the target. Rectangular collimator slits defined an acceptance angle of 0.8$^\circ$. The measured angular range extended from 14$^\circ$ to 160$^\circ$ (laboratory frame). 
These telescopes allow to discriminate ions with very different ranges. For instance, low-energy protons originated from the deuteron breakup, and tritons produced by the neutron-pickup reactions, were identified in the $\Delta E_1$ vs. $E_{\rm{tot}}$ spectrum, Fig. \ref{spectrum-2D} left ($E_{\rm{tot}}$ is the sum of the signals of the four detectors). In some cases, the elastic scattering peak was better resolved in the $\Delta E_1 + \Delta E_2$ vs.  $E_{\rm{tot}}$ spectrum, Fig. \ref{spectrum-2D} right. High energy protons, produced by the neutron-stripping reactions, were also discriminated by the $\Delta E_1 + \Delta E_2$ vs. $E_{\rm{tot}}$ spectrum. The elastically scattered deuterons were selected by appropriate gating in the 2-dimensional spectra.

\begin{figure}[h!t]
  \centering
  \subcaptionbox{\label{fig:Etot-todo-1}}
  {\includegraphics[width=0.75\textwidth]
  {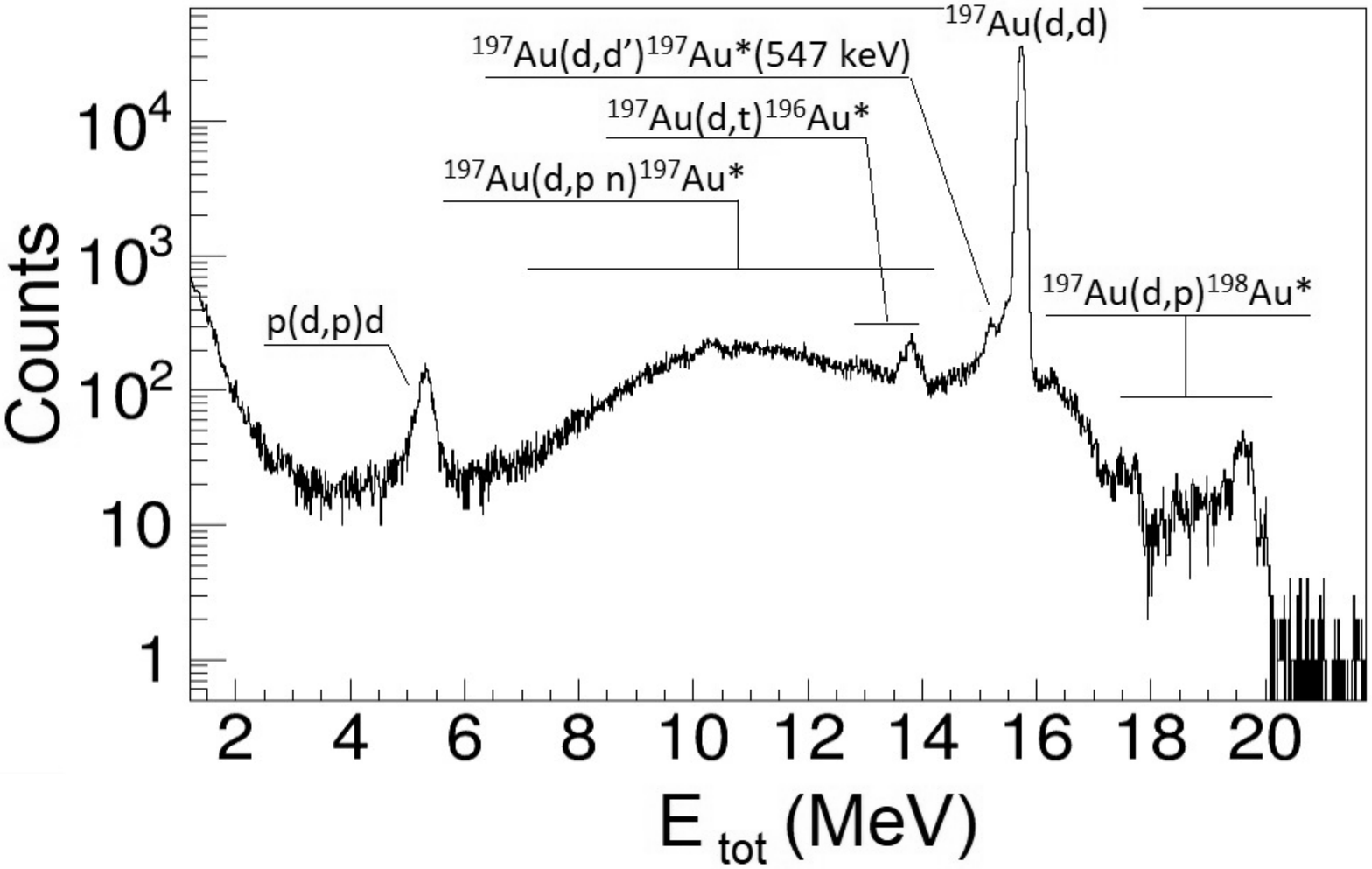}}
  \subcaptionbox{\label{fig:Etot-todo-2}}
  {\includegraphics[width=0.75\textwidth]
  {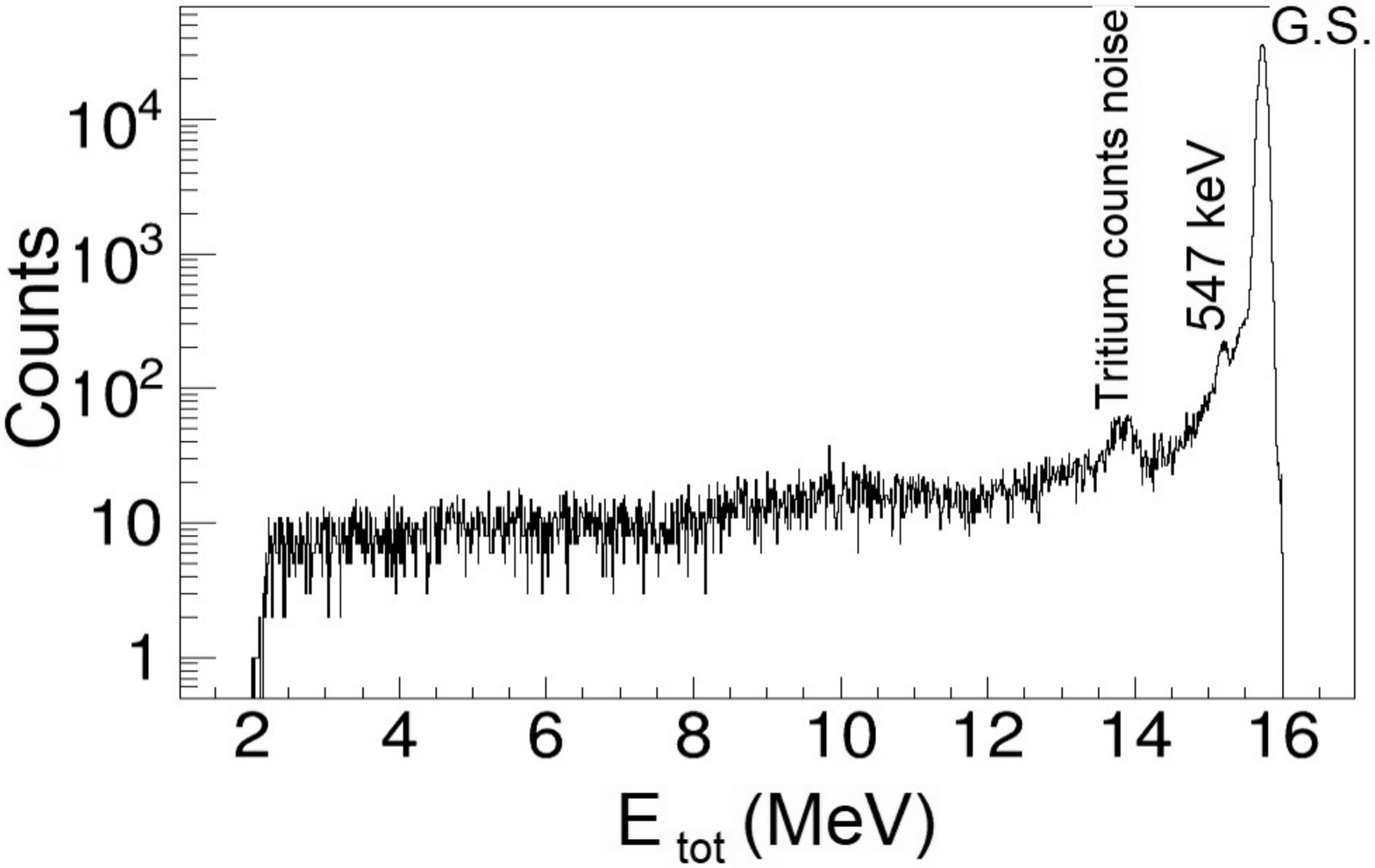}}
  \caption{Typical total energy spectrum (sum of the four detector signals) recorded at $\theta_{\rm lab} = 53^\circ$ for $E_{\rm lab} = 16$~MeV. Upper panel \ref{fig:Etot-todo-1}) shows the spectra without applying any gate on it. The wide bump to the left of the elastic peak is mainly due to proton breakup; other peaks signal are inelastic scattering, tritium from the pickup process, or possible not identified contaminants. At energies above 16 MeV (the elastic peak) and up to 20.3  MeV (beam energy plus the Q-value energy), the neutron-stripping (d,p) peaks corresponding to various excited states of $^{198}$Au were found. At the lower panel \ref{fig:Etot-todo-2}), the whole spectrum $^{198}$Au was filtered with a deuteron gate. The prominent peak corresponds to elastic scattering, whereas the small ones to the left to inelastic scattering. Referenced $^{197}$Au excitation energy level is marked.}
\label{spectrum-1D}
\end{figure}

With a second gate, a clean spectrum of protons can be extracted. This spectrum includes higher energy protons stemming from the neutron stripping and the lower energy ones produced by the breakup process. A third gate selects tritons ejected from the neutron pickup reaction. Reactions leading to protons and tritons will be presented elsewhere. A typical 1-dimensional spectrum ($\theta_{\rm lab} = 53^\circ$, $E_{\rm lab} = 16$~MeV) is shown in Fig. \ref{spectrum-1D}. An asymmetric Gaussian curve was fitted to the histograms for the peak integration.

Two normalization methods were applied. The first one used a Faraday cup at the end of the beamline, far away from the target, which integrated the total charge delivered by the beam during each run. The drawback of this method is that it relies on the geometrical measurement of solid angles and the target thickness. These quantities are difficult to be measured with low uncertainty values. A more reliable normalization was obtained using two simple silicon detectors (monitors) placed at fixed angles of $\pm 16^\circ$ from the beam direction, where the elastic scattering is supposed to be pure Rutherford. Relative uncertainties typically ranged from 2\% to 4\%, except at the higher energies and backward angles, where they reached values of 10\% or 14\% due to the low counting recorded spectra.

\section{Theoretical analysis} \label{sec:elastic}

\subsection{Critical and absorption distances} \label{sec.dids}
The absorptive processes induced by peripheral reactions, like transfer,  occur at grazing distances. In the OM, these processes are accounted mainly by the imaginary component of the potential, which affects the angular distribution of the elastic scattering. It has been shown (Ref. \cite{GAA20} and refs. therein)  that the elastic scattering is mainly sensitive to the potential in a very narrow radial distance around the so-called sensitivity radius $R_S$. A hint to this distance can be obtained by the semiclassical approach of \cite{GKA16,PR04}. In those works, the reduced critical interaction distance $d_I$ and the reduced strong-absorption distance $d_S$ are defined as those for which $d\sigma_{el}/d\sigma_{\rm {Ruth}}$ equal to 0.98 and 0.25, respectively. From now on, $\sigma/\sigma_{\rm {Ruth}}$ is going to be used instead $d\sigma_{el}/d\sigma_{\rm {Ruth}}$. The reduced distance is  $d=D/(A_p^{1/3}+A_T^{1/3})$, with being $D$ the distance of closest approach. These values can be easily obtained from the experimental elastic cross-section data. 

\begin{figure}[h!t]
  \centering
  \includegraphics[width=0.85\linewidth]{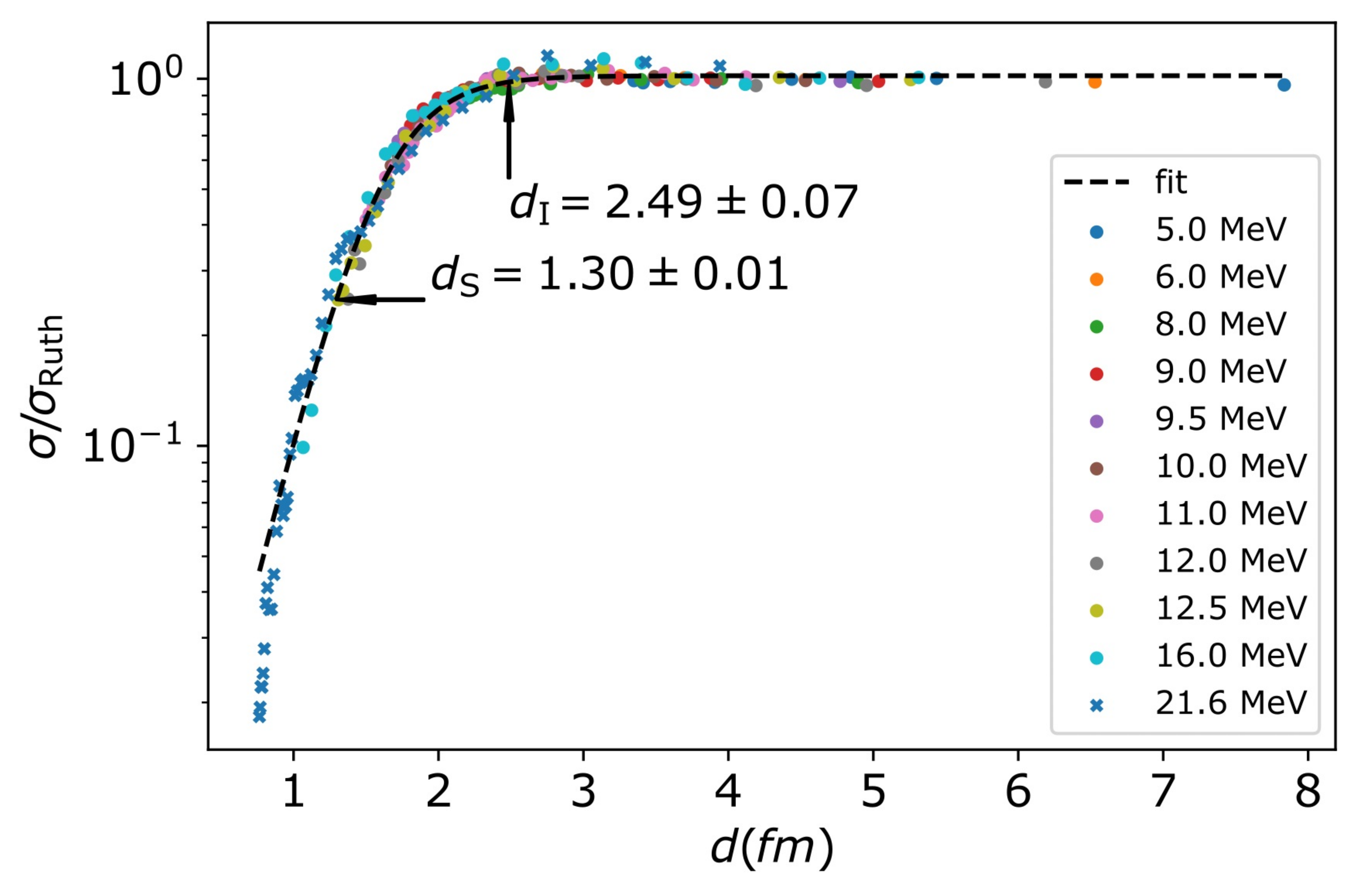}
  \caption{(color online) The ratio of the elastic scattering to Rutherford cross section versus the reduced distance of closest approach. The dashed curve is the parametrization $\sigma/\sigma_{\rm {Ruth}}=p_1[1+e^{d_1(d-d_2)}]^{-1}$ with $p_1=1.020$, $d_1=-3.628$ and $d_2=1.608$.  Solid circles represent data obtained from the present work, while crosses represent data from Ref. \cite{datos21_6mev}.}
\label{fig.dmax}
\end{figure}

Figure \ref{fig.dmax} shows, in a semi-logarithmic scale, the  elastic cross section $\sigma/\sigma_{\rm {Ruth}}$ as a function of $d$ for all deuteron bombarding energy measured in the present work. Additional data for the energy of 21.6 MeV, taken from Ref. \cite{datos21_6mev}, were also included.

By using the Boltzmann-type exponential function of Ref. \cite{GKA16}
\begin{equation}
   \frac{ \sigma}{\sigma_{\rm {Ruth}}}
       =\frac{p_1}{1+e^{d_1(d-d_2)}} \, ,
\end{equation}
the parameters $p_1$, $d_1$ and $d_2$ are optimized from the experimental elastic scattering cross section. The fitting curve using the adopted values of these parameters is shown in Fig.~\ref{fig.dmax}. With the mentioned above prescription the value $d_I=2.49$~fm and $d_S=1.30$~fm were obtained. Comparing these results to those of other projectile-target systems allows us to get an insight into the nuclei main features, whether they are strongly or weakly bound (see Table \ref{table.rr}). Our calculated $d_I$ for the $\rm{d} + ^{197}\rm{Au}$ system is very similar to the ones of weakly bound projectiles, probably due to the extension of the deuteron wave function, as a consequence of the dominance of the $s$ partial wave in it. The separation energy ($E_{\rm{sep}}$), i.e., the lowest energy needed to separate a nucleus into a particular cluster configuration, is given for comparison.
\begin{table}[h!t]
\caption {\label{table.rr} Calculated reduced critical interaction $d_I$ and the reduced strong-absorption  $d_S$  distances compared with those calculated in Ref. \cite{GLK18} for some strongly and weakly bound projectiles.}
\centering
\begin{tabular}{c|cccc}
\hline
 Ref.           & System  &     $E_{\rm{sep}}$  (MeV)       &  $d_I$ (fm) & $d_S$ (fm) \\
\hline
 this work      & $d+^{197}$Au    &  2.224    & 2.49(7)  & 1.30(2) \\
\hline
    M. Cubero 12 \cite{cubero} & $^{11}$Li+$^{208}$Pb & 0.369 & 5.2(4) & 1.59(4) \\
    S.-Benitez 05 \cite{S05}  & $^{6}$He$+^{208}$Pb & 0.973  & 2.20(5)  & 1.589(7) \\
 M. Duran 16 \cite{M16} & $^{8}$He$+^{208}$Pb & 2.140  & 2.24(7)  & 1.718(6) \\
 Keeley-94 \cite{K94}  & $^{6}$Li$+^{208}$Pb & 1.474 & 1.95(4) & 1.521(5)   \\
 Keeley-94 \cite{K94} & $^{7}$Li$+^{208}$Pb &  2.467 & 1.74(2)  & 1.491(3)   \\
 Yu-10 \cite{Y10} & $^{9}$Be$+^{208}$Pb & 1.655 & 1.86(2)  & 1.540(4)   \\
 \hline
 Santra-01 \cite{ssk01} & $^{12}$C$+^{208}$Pb & 7.275 & 1.66(1)  & 1.491(2) \\
 Vulgaris 86 \cite{vgs86}  & $^{16}$O$+^{208}$Pb & 7.162 & 1.64(1)  & 1.498(2)  \\
\hline
\end{tabular}
\end{table}

In our case, the critical distance gives $17.6$~fm and the strong-absorption $9.2$~fm. These values would indicate that fusion will take place for approximation distances smaller than $9.2$~fm. For distances greater than $17.6$~fm, the elastic scattering would be mostly Rutherford, and the absorption due to nuclear effects would lie between these two values. In Sec. \ref{sec.sr} it will be shown that the value for the imaginary sensitive radius, the distance at which the elastic scattering is most sensitive, is $R_S=11.5$~fm.

\subsection{Optical potentials}
The phenomenological Woods-Saxon potential (WSP) and the semi-microscopic S\~ao Paulo potential (SPP) are used in the theoretical analysis to address the independence of the conclusions about the deuteron threshold property. These potentials take the form:
\begin{align}
    U(r) &= V_C(r) - V_0 f(r,r_0,a) 
        + i\, 4 a_s W_s \frac{df}{dr}(r,r_{s0},a_{s})\, , \label{potWS}  \\
    V_{SPP}(r,E) &= V_C(r)+ \left[ 
                  N_R(E) + i N_I(E) \right]
                  V_{LE}(r,E) \, , \label{potSP}  
\end{align}
being $V_C$ the Coulomb interaction for a uniformly charged distribution. The function $f(r)=[1+exp((r-R)/a)]^{-1}$ refers to the Fermi distribution, with the system radius given by the usual formula $R=r_0(A_p^{1/3}+A_T^{1/3})$ and diffuseness $a$. The energy-dependent radial form factor $V_{LE}(r, E)$ represents the local equivalent double-folding S\~ao Paulo potential.

\begin{figure}[h!t]
\centering
 \centering
 \includegraphics[width=0.85\textwidth]{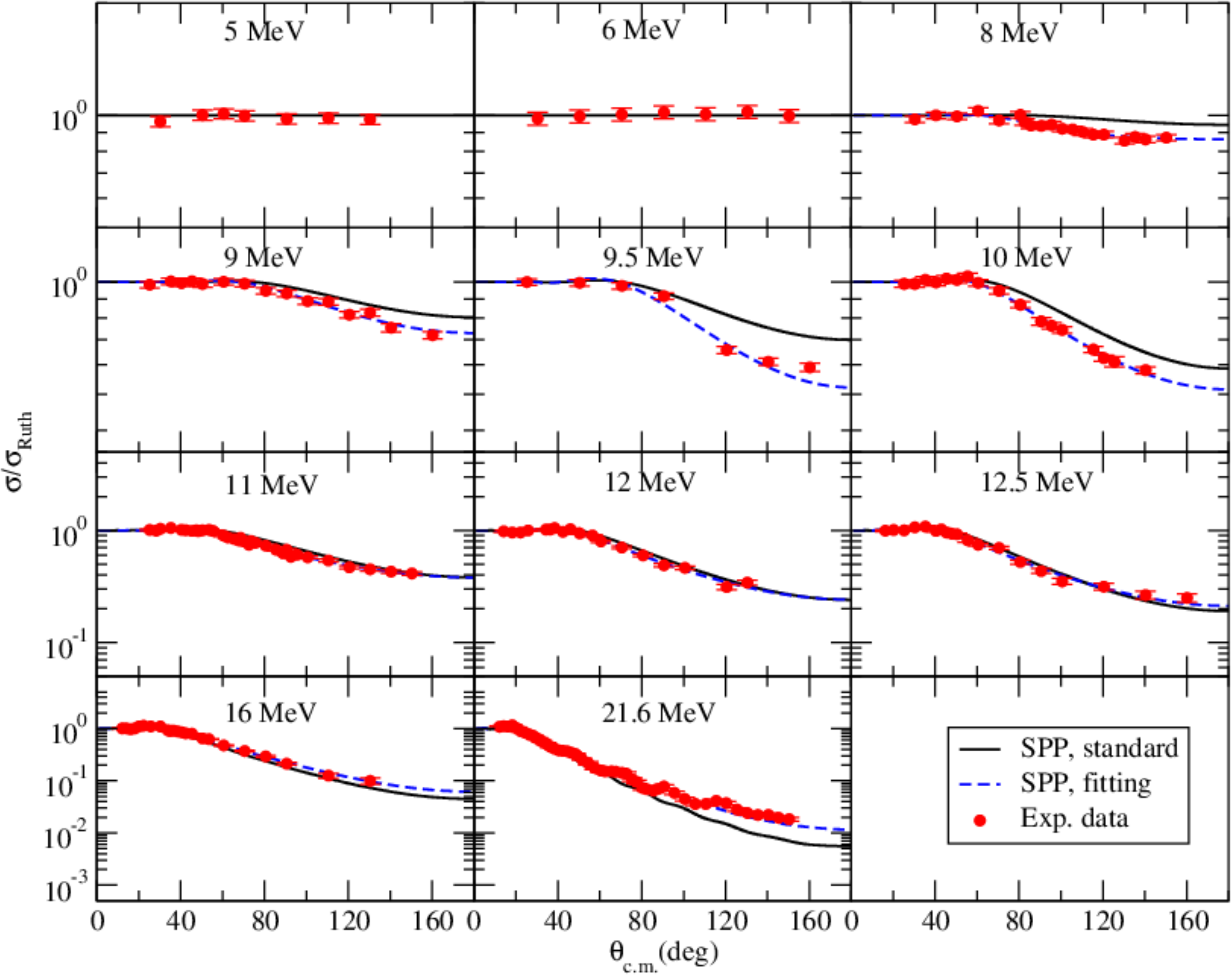}
\caption{(color online) Experimental elastic cross sections (normalized to Rutherford) compared with the calculated ones using the S\~ao Paulo potential. The black solid line uses  the standard prescription $N_R =1$ and $N_I=0.78$, whereas the blue dashed line shows the best fitting values for $N_R$ and $N_I$ from $\chi^2$ minimization. Note that the angular distributions for energies of 5 and 6 MeV are indistinguishable from Rutherford and, hence, will not be analyzed hereafter.} 
\label{fig:elastic}
\end{figure}

Figure \ref{fig:elastic} shows the experimental elastic cross section $\sigma/\sigma_{\rm {Ruth}}$ and the calculated one using SPP from the systematic parametrization of Ref. \cite{AFL19} $N_R=1$ and $N_I=0.78$ (labeled as SPP, standard); one can see that the agreement with the data is poor. Using the code SFRESCO \cite{T88}, the real $N_R$ and imaginary depths $N_I$ were simultaneously optimized for each energy . The results are represented by the curve labeled SPP fitting. 
The figure shows an improvement in the adjustments for most the bombarding energies. Table \ref{tab:chiSP} shows the behavior of $N_R$ and $N_I$ as a function of energy. The values are not as smooth as expected, but a remarkable increase of $N_I$ for $8$ MeV is found ($N_I = 32.1$, more than one order of magnitude relative to the higher energy corresponding values). Similar behavior was found for the phenomenological WSP (as can be seen comparing the values from Table \ref{tab:var2}).
\begin{table}[h!t]
\caption{\label{tab:chiSP} Optimized SPP parameters and $\chi^{2}$ with respect to the number of data points $N$.}
\centering
\begin{tabular}{cccc}
\hline
 {E (MeV)}
 &${N_R}$
 &${N_I}$
 & {$\frac{\chi^{2}}{ N}$}  \\ 
\hline
21.6 & 1.0  & 1.4 & 2.97 \\ 
16 & 0.56 & 0.65  & 3.58 \\ 
12.5 & 1.7 & 2.09 & 1.32 \\ 
12 & 1.86 & 2.03 & 1.76 \\ 
11 & 2.36 & 3.80 & 1.71 \\ 
10 & 2.37 & 2.71 & 0.37 \\
9.5 & 3.11 & 0.74  & 0.92 \\
9 & 2.25 & 2.5  & 0.94 \\
8 & 2.27 & 32.1 & 0.96 \\
\hline
\end{tabular}
\end{table}

The optimization of the WSP parameters is done iteratively, starting from energy independent parameters. First, the geometrical parameters obtained from Ref. \cite{supersymmetry} were taken and kept fixed: $r_0$=1.16, $r_{s0}$=1.21, $a$=0.81, and $a_{s}$=0.67 fm, while  the real and imaginary depths, $V_0$ and $W_{s}$, were adjusted to the deuteron elastic cross section at 21.6 MeV \cite{datos21_6mev}. The optimization was carried out using the code SFRESCO, yielding  $V_0$=101 MeV and  $W_{s}=29.1~\rm{MeV}$; with $\chi^2/N=1.58$, where $N$ is the number of data points. This case was taken as a reference because a large number of experimental data give a starting point for the depths optimizing procedure for each bombarding energy. Throughout the calculation of the elastic scattering cross section using this set of parameters for each deuteron incident energy and comparing it with the experimental measurement data, the obtained $\chi^2/N$ values shown in Table \ref{tab:varE} were obtained . The poor quality of the adjustments (large $\chi^2/N$ values) at the lower energies reflects the impossibility to describe the threshold anomaly with an energy-independent potential. 

\begin{table}[h!t]
\caption {\label{tab:varE} Energy independent Woods-Saxon potentials: $V_0$=101, $W_{s}$=29.1 MeV; $r_0$=1.16, $r_{s0}$=1.21, $a$=0.81, $a_{s}$=0.67 fm. The  global estimator given by Eq. (\ref{estimator}) is $\chi_{G}^{2}$=13.08.\\
\hspace{1cm}}
\centering
\begin{tabular}{lccccccccc}
\hline
 E (MeV)& 21.6 & 16 & 12.5 & 12 & 11 & 10 & 9.5 & 9 & 8  \\ \hline
\quad 
$\chi^{2}/N$ & 1.58  &  4.68  &  6.91  &  11.24  &  33.68  &  29.12  &  0.84  &  11.57  &  15.17 \\ \hline
\end{tabular}
\end{table}

Let us introduce a global estimator parameter $\chi_{G}^{2}$ in order to encompass all calculated $\chi_i^{2}/N_i$ values for each incident energy $E_i$,
\begin{equation}
\label{estimator}
    \chi_{G}^{2} = \frac{\sum^{n}_{i=1} 
        (\chi_{i}^{2} / N_{i})  N_{i} }{(\sum^{n}_{i=1} N_{i})-p} 
\end{equation}
with $p$ the number of optimized parameters, $N_i$ the number of data points for the $i^{th}$ energy, and $n=9$ the total number of measured energies (including the 21.6 MeV energy  referenced data).

Taken the geometrical parameters of Table \ref{tab:varE}, and the depth $V_0$=101, $W_{s}$=29.1 MeV as initial guess, the WSP depths separately for each deuteron incident energy were optimized, and the results are shown in Table \ref{tab:var2}. The fitting method strongly improves at the barrier energy regime, where the threshold anomaly occurs. The global estimator gives $\chi_{G}^{2}=1.57$ greatly improving the previous one since all energies were now independently adjusted.  

\begin{table}[h!t]
\caption {\label{tab:var2} Optimized Woods-Saxon potentials depths $V_0$ and $W_{s}$ for fixed geometry ($r_0$=1.16, $r_{s0}$=1.21, $a$=0.81, $a_{s}$=0.67 fm). The calculated global estimator is $\chi_{G}^{2}=1.57$.\\
\hspace{1cm}}
\centering
\begin{tabular}{c|ccc}
\hline
 \multirow{2}{*}{E (MeV)} & Volume WSP & Surface WSP & \multirow{2}{*}{$\frac{\chi^{2}}{ N}$}  \\ 
\cline{2-3} 
 & $V_0$ (MeV)  &  $W_{s}$ (MeV) &\\  
 \hline
21.6  &  101.2  &  29.2  &  1.58\\
16  &  78.5  &  24.1  &  3.40\\
12.5  &  202.8  &  113.3  &  1.22\\
12  &  216.4  &  112.5  &  1.82\\
11  &  250.8  &  211.6  &  1.69\\
10  &  251.0  &  159.3  &  0.35\\
9.5  &  233.7  &  129.2  &  0.30\\
9  &  173.8  &  149.4  &  1.66\\
8  &  69.6  &  1695.2  &  0.59\\
\end{tabular}
\end{table}

To further improve the adjustments, the geometry parameters were allowed to change. Firstly, the reduced radii were optimized together with the depths. The global estimator yielded $\chi_G=1.26$, further improving the previous value. Table \ref{tab:var4} shows the result. The magnitude of the depths changed substantially with respect to the values quoted in Table \ref{tab:var2}. The behavior of the reduced radii, as a function of energy for both volume and surface interaction, mainly increases as the energy decreases.

\begin{table}[h!t]
\caption {\label{tab:var4} Optimized values of the depth and reduced radii, for $a$=0.81, $a_{s}$=0.67 fm. The deduced global estimator is $\chi_{G}^{2}$=1.26.\\
\hspace{1cm}}
\centering
\centering
\begin{tabular}{c|ccccc}
\hline
 \multirow{2}{*}{E [MeV]}&\multicolumn{2}{c}{Volume WSP}&\multicolumn{2}{c}{Surface WSP} & \multirow{2}{*}{$\frac{\chi^{2}}{ N}$}  \\ 
\cline{2-5} 
 & $V_0$ (MeV)  & $r_0$ (fm)  & $W_{s}$ (MeV) & $r_{s0}$ (fm) & \\
 \hline
21.6 & 95.5 & 1.20 & 23.2 & 1.27 & 0.88\\
16 & 78.7 & 1.16 & 16.5 & 1.27 & 3.01\\
12.5 & 67.8 & 1.30 & 17.2 & 1.43 & 1.15\\
12 & 67.7 & 1.31 & 19.2 & 1.41 & 1.72\\
11 & 86.5 & 1.29 & 16.5 & 1.50 & 1.52\\
10 & 64.5 & 1.34 & 21.7 & 1.44 & 0.36\\
9.5 & 39.9 & 1.35 & 15.5 & 1.45 & 0.09\\
9 & 39.3 & 1.33 & 7.5 & 1.58 & 0.86\\
8 & 6.3 & 1.36 & 57.6 & 1.60 & 0.59\\
\end{tabular}
\end{table}

\begin{figure}[h!t]
 \centering
 \includegraphics[width=0.85\textwidth]{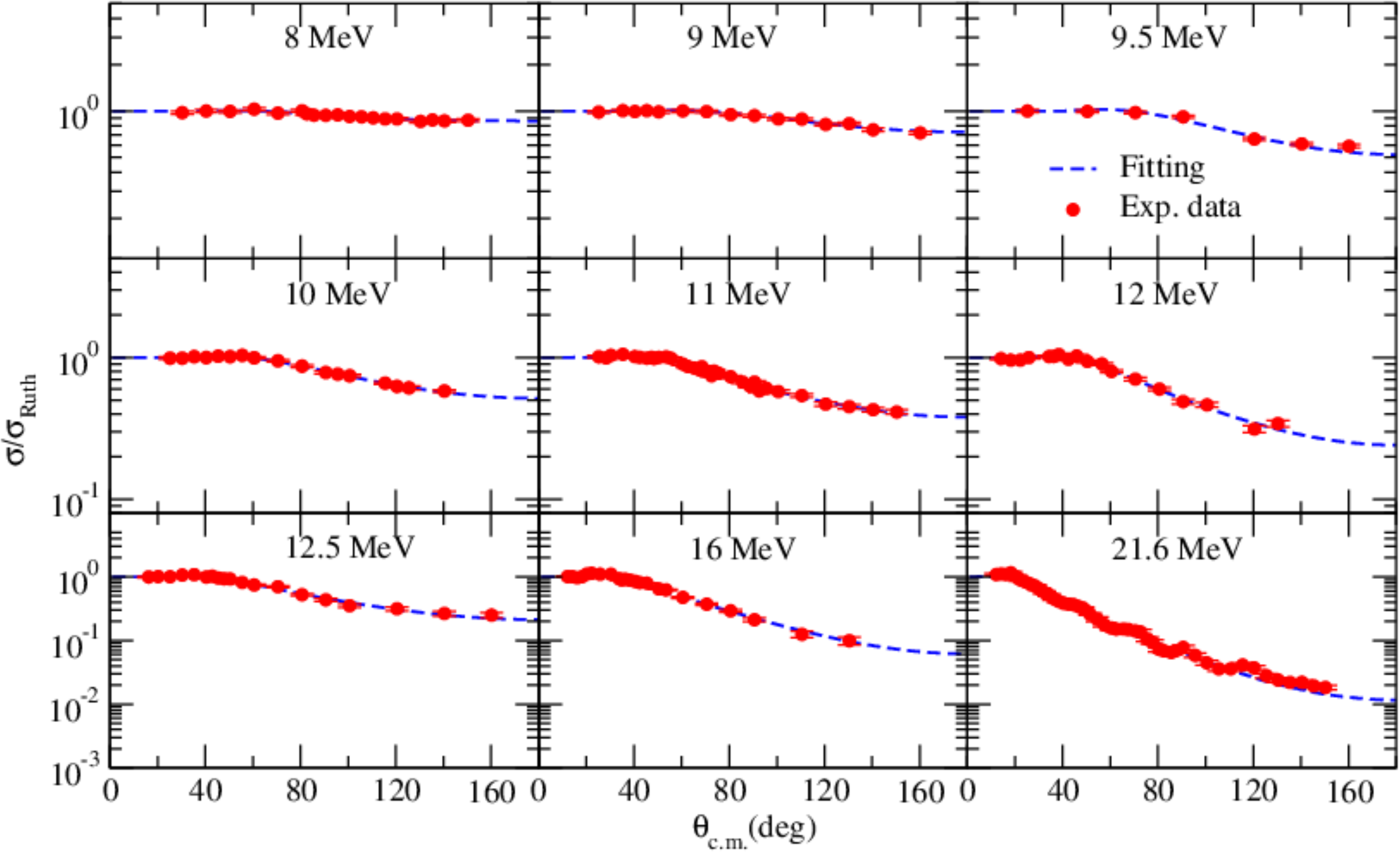}
\caption{Comparison of the experimental elastic cross sections with the calculated one using the WSP parameters of Table \ref{tab:varTODO}.}
\label{fig:elasticWS}
\end{figure}

Secondly, all parameters are allowed to change during the $\chi^2$ optimization. Table \ref{tab:varTODO} shows the different parameters for each energy. The new optimized reduced radii are consistent with previous optimization and the potential depths. Comparison of the $\chi^2/N$ values with those of Table \ref{tab:chiSP} shows that the values for the WSP are smaller since it has many more parameters to be fitted. It is still remarkable that in the case of SPP, similar goodness of the fits is attained using only two parameters, which is more realistic from a physical point of view. The global estimator for the WSP improved to the value $\chi_{G}^{2}$=1.18; as for the SPP potential, a value of $\chi_{G}^{2}$=1.97 was reached. The calculated cross section with this set of WSP parameters is compared with the measured one in Fig. \ref{fig:elasticWS}. 

\begin{table}[h!t]
\caption{\label{tab:varTODO} Optimized WSP parameters.}
\centering
\begin{tabular}{c|ccccccc}
\hline
 \multirow{2}{*}{E (MeV)}
 &\multicolumn{3}{c}{Volume WSP}
 &\multicolumn{3}{c}{Surface WSP} 
 & \multirow{2}{*}{$\frac{\chi^{2}}{ N}$}  \\ 
\cline{2-7} 
 & $V_0$ (MeV)  & $r_0$ (fm) & $a$ (fm) & $W_{s}$ (MeV) & $r_{s0}$ (fm) & $a_{s}$ (fm) & \\
 \hline
21.6 & 108.7 & 1.10 & 0.89 & 19.3 & 1.28 & 0.73 & 0.78 \\ 
16 & 61.2 & 1.11 & 0.86 & 11.4 & 1.25 & 0.80 & 2.90 \\ 
12.5 & 66.3 & 1.30 & 0.85 & 24.4 & 1.43 & 0.61 & 1.11  \\ 
12 & 65.7 & 1.34 & 0.75 & 15.4 & 1.40 & 0.73 & 1.63  \\ 
11 & 60.9 & 1.38 & 0.77 & 12.6 & 1.55 & 0.64 & 1.40  \\ 
10 & 50.3 & 1.34 & 0.90 & 36.8 & 1.44 & 0.60 & 0.30  \\
9.5 & 73.3 & 1.35 & 0.73 & 14.4 & 1.45 & 0.64 & 0.08  \\
9 & 38.9 & 1.33 & 0.75 & 5.2 & 1.58 & 0.75 & 0.68  \\
8 & 90.0 & 1.36 & 0.77 & 91.8 & 1.60 & 0.62 & 0.58  \\
\hline
\end{tabular}
\end{table}

\section{Sensitivity radius} \label{sec.sr}
\begin{figure}[h!t]
\centering
\subcaptionbox{\label{notch}}
{\includegraphics[width=0.49\textwidth]{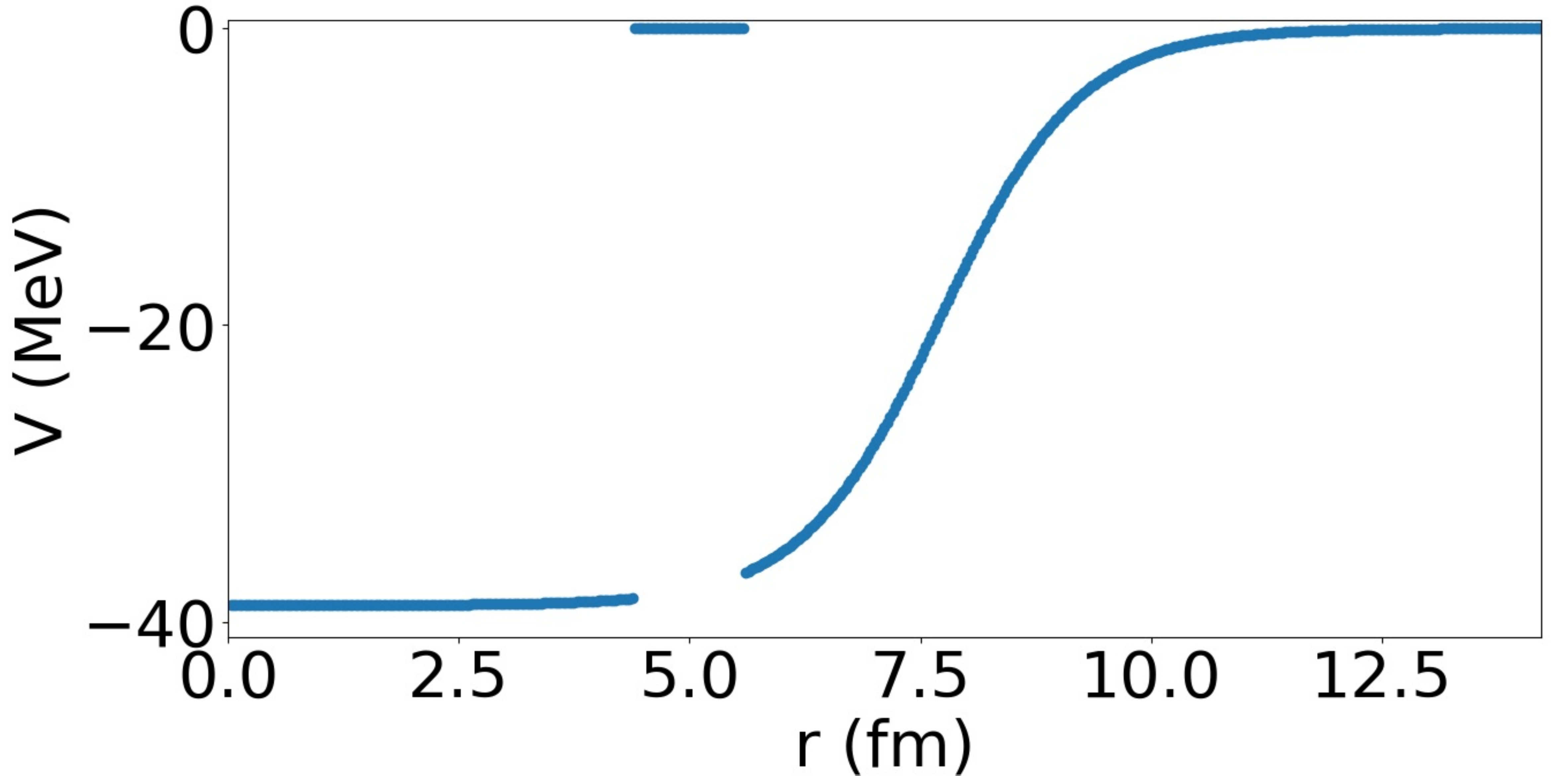}}
\subcaptionbox{\label{chi2}}
{\includegraphics[width=0.49\textwidth]{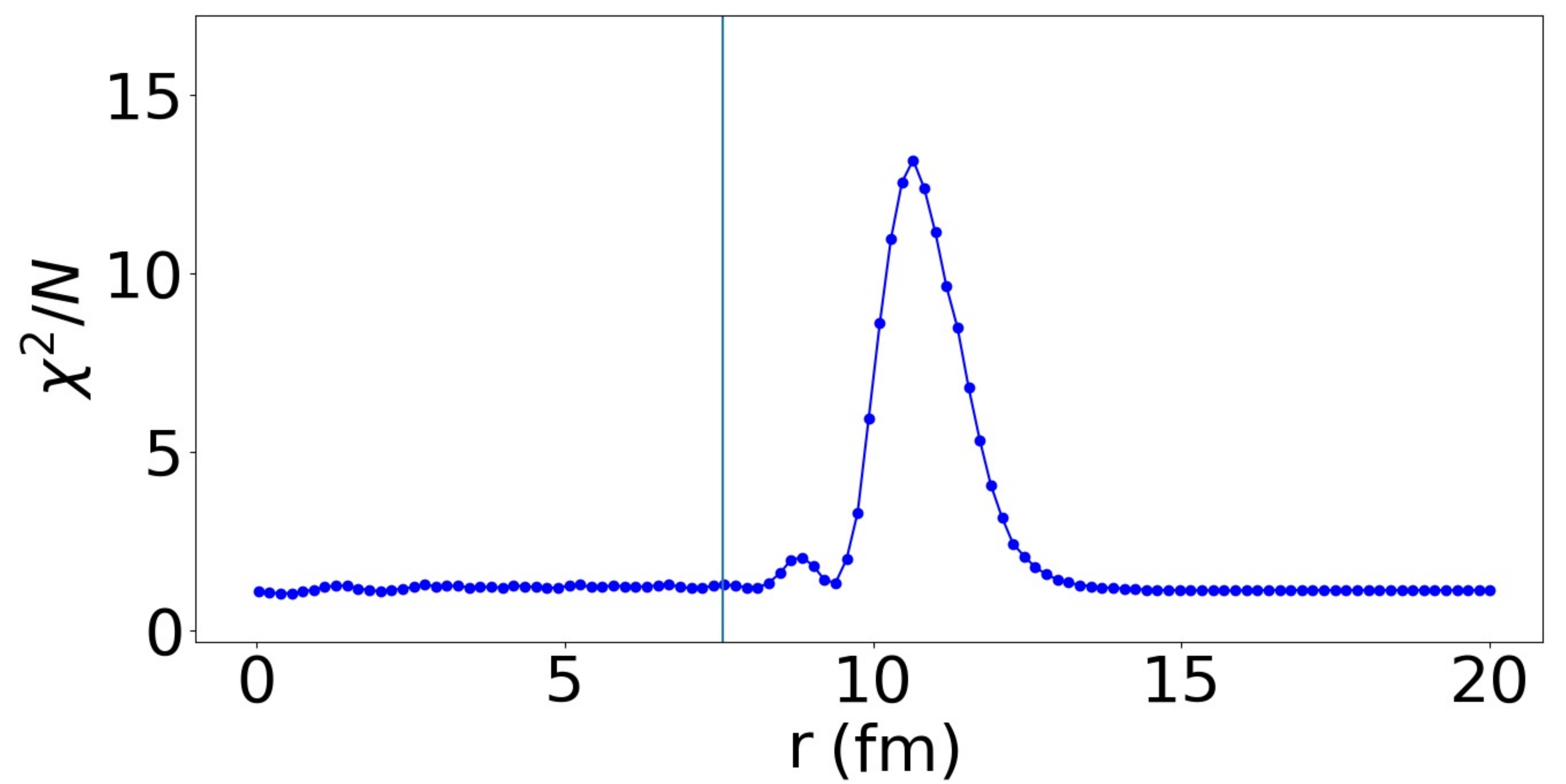}}
\caption{ (a) Example of a notched potential for the real part of the nuclear potential of Eq. (\ref{potWS}). A discontinuity can be seen at $r=5$ fm. (b) $\chi^2/N$ versus the notch's position. The sensitivity radius is defined as the rightmost maximum of $\chi^2/N$ value. The vertical line gives the position of $R$.}
\label{notch-chi2}
\end{figure}

The optimization analysis described in the previous section shows the ambiguity of parameters setting of the optical potential, i.e., different sets of depths and geometrical potential parameters reproduce the elastic scattering cross section equally well \cite{i59}. This property of the optical potential is a consequence of the fact that elastic scattering only probes the surface of the projectile-target interaction reaction, giving rise to the concept of sensitivity radius $R_S$ \cite{fbl85,s91}. As an alternative the usual parameters variation \cite{s91,VRM21} for the determination of the sensitivity radius, the so called notch method was used \cite{cd80}. This last method is somehow more intuitive since one analyzes the effect on the cross-section addressed by a localized perturbation on the radial form factor of the potential, as shown in Fig. \ref{notch-chi2}.

The sensitivity radius $R_S$ is defined as the rightmost maximum of $\chi^2/N$ versus $r$, being $r$ the position of the notch position-center. The notch length was taken to be $1.2$ fm because this width gives a clear maximum for all the  energies for both the real and imaginary components of the optical potential. The average sensitivity radii obtained for the real and imaginary parts of the optical potential are $10.5$ and $11.5$ fm, respectively; then, the following value was taken for $R_S=11.0 \pm 0.5$ fm. This figure is much greater than for $^7$Li on $^{208}$Pb, $R_S=8.2 \pm 0.4$ fm \cite{VRM21}, but smaller than for $^9$Be on $^{197}$Au, $R_S=12.1 \pm 0.2$ fm \cite{GAA20}. These results suggest that some connection between the value of $R_S$ and the breakup threshold energy might exist.

One possibility is to use the $R_S$ value to calculate the OM potential precisely at that point and, in doing so, enable the possibility of comparing OM potentials of very different shapes. In the next section, the radial integral of the OM potentials in the neighborhood of the sensitivity radius are calculated as a function of energy.

\section{Threshold anomaly analysis} \label{sec.ta}
A magnitude that isolates the information given by the optical potential at the sensitivity radius is the Gaussian weighting integral $[G(E)]_U$ \cite{bam93}, 
\begin{equation}\label{eq.GE}
  [G(E)]_U = \frac{4 \pi}{A_p \times A_t}
    \int^{\infty}_0 G(r,R_S,\sigma) U(r,E) r^2 dr \, ,
\end{equation}
where $U(r,E)$ represents the real or imaginary components of the WSP or SPP optical potentials for a given incident energy of the projectile; and $G(r,R_S,\sigma)$ the normal or Gaussian distribution \cite{dataanalysis}, centered at $R_S=11.0$ fm, with $\sigma=0.5$ fm.
 
\begin{figure}[h!t]
\centering
\subcaptionbox{\label{RealInts}}
{\includegraphics[width=0.75\textwidth]{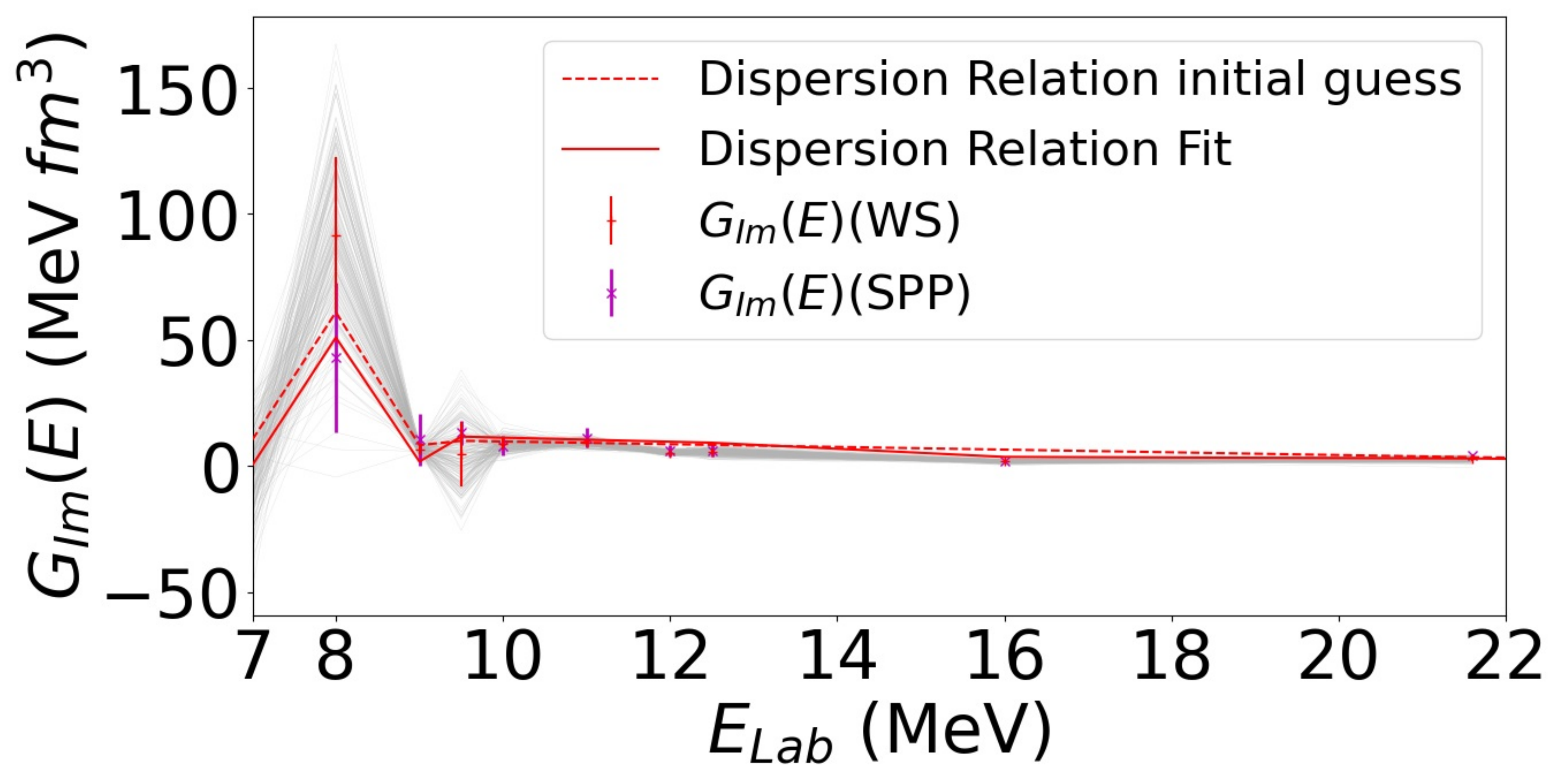}}
\subcaptionbox{\label{ImagInts}}
{\includegraphics[width=0.75\textwidth]{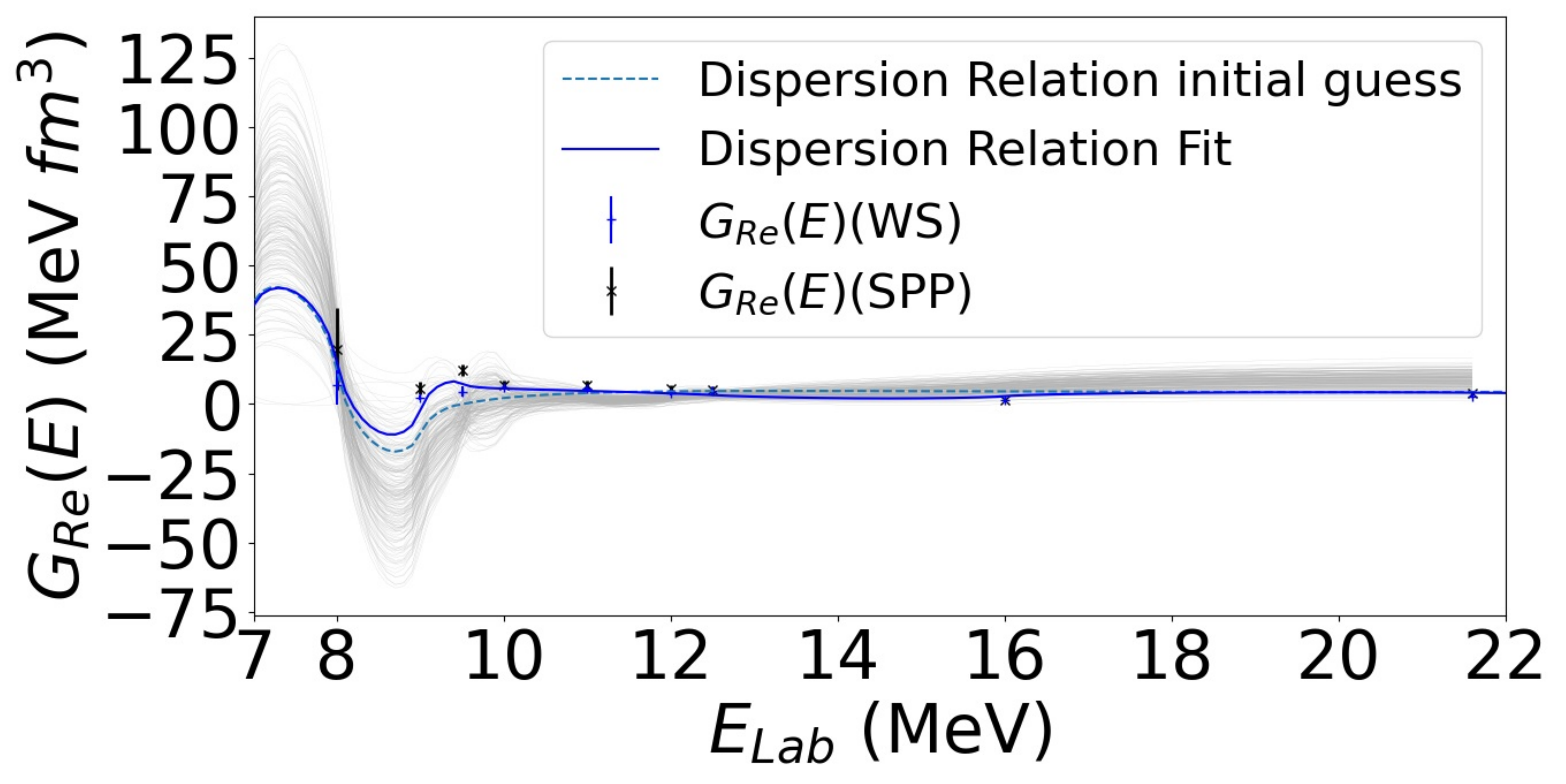}}
\caption{ Gaussian weighting integral $[G(E)]_U$ calculated as explained in the text. The parameters of Table \ref{tab:varTODO} were used for the WS potential. (a) $[G(E)]_U$ with $U$ the imaginary part of the WSP and SPP. (b) idem (a) for the real part of the optical potential. The light grey lines represent Gaussian variations of the curves labeled Dispersion Relation Fit.}
\label{Integrals}
\end{figure}

Figures \ref{Integrals}(a) and (b) show the calculated weighting integral using Eq. (\ref{eq.GE}) for the imaginary and real parts of the optical potentials, respectively. The SPP and WSP optimized parameters of Tables \ref{tab:chiSP} and \ref{tab:varTODO} were used for these calculations. One can observe a very good agreement between both optical potentials, even when very different parametrizations rule them. The uncertainties were calculated considering the contributions from the optical potential and the variation produced when changing $\sigma=0.125$ fm. The error bars substantially increase as the energy decreases below the Coulomb barrier ($V_{B}=10.5$ MeV).

It is well established that causality entangles the real and imaginary components of the optical potential \cite{mns86}. The same relation is valid for the volume integral $[G(E)]_U$,
\begin{equation}\label{eq.dr}
    \Delta [G(E)]_{Re} = C +  \frac{1}{ \pi} \textbf{P}\int^{\infty}_{-\infty} 
            \frac{[G(E)]_{Im} }{(E-E')} dE' \, ,
\end{equation} 
with $C$ a constant, $\textbf{P}\int$ the Cauchy’s principal value, and $[G(E)]_{U}$ the integrals for the real or imaginary part of the potential $U$. A piecewise initial guess is made for the imaginary component of the volume integrals \cite{s91}, smoothly connecting the values calculated using Eq. (\ref{eq.GE}) for the Woods-Saxon potential, as can be seen in Fig. \ref{Integrals}(a) (dashed curve). The zero value for the curve Dispersion Relation Fit was taken around 7 MeV since, at this energy, the scattering is pure Coulomb and utterly insensitive to the nuclear potential. We calculated for different values of the drop for the imaginary part to zero and found that it slightly changed the real part of the integral without changing the physical conclusions. Then, $\Delta [G(E)]_{Re}$ is calculated from Eq.(\ref{eq.dr}), and shown in Figure \ref{Integrals}(b). The constant $C$ was fixed by minimizing the squared differences between the calculated volume integrals using Eq. (\ref{eq.GE}) for the real part data. The agreement of $\Delta [G(E)]_{Re}$ with the values calculated using Eq. (\ref{eq.GE}) is poor. The calculated $\chi^2$ for the imaginary and real parts are 12.6 and 38.0, respectively. These figures may be improved by simutaneously minimizing the imaginary and real curves (the Iminuit \cite{iminuit} Python package was used as minimizer.) The optimization procedure gives $\chi^2$=14.2 and 7.3 for the real and imaginary parts, respectively. This pair of curves are shown in Fig. \ref{Integrals}(a) by the continuous red line, and its counterpart calculated using the dispersion relation is indicated by a continuous blue line in Fig. \ref{Integrals}(b). It can be seen that the agreement was improved.

The real part of the volume integral does not show the characteristic bell behavior of the threshold anomaly, which seems to indicate that the deuteron, at least incoming onto  $^{197}$Au, has a breakup anomaly behavior energy-dependence of the optical potential. To better assess the absence of the characteristic bump around the coulomb barrier in the real integral, we made a $\chi^2$ analysis. We generated two hundred imaginary curves with gaussian probability, and then we calculated the real counterpart from the dispersion relation equation. Figure \ref{Integrals}(b) shows a remarkable concentration of curves following the fitted curve, confirming the absence of usual threshold anomaly.

\section{Discussion and conclusions}\label{sec:discussion}
Differential elastic cross sections for the $^2$H+$^{197}$Au reaction were measured at ten incident energies between 5 and 16 MeV. Although transfer reaction products had been detected down to 5 MeV, the elastic cross section obtained for 5 and 6 MeV resulted indistinguishable from Rutherford scattering.

A closest approach distance analysis was performed, yielding a value of  $d_I=2.49(7)$ fm for the reduced critical interaction distance. This value  is closer to those of the borromean halo $^6$He and $^8$He than to the other stable weakly bound nuclei. 

Optical WSP and SPP were optimized to describe the experimental data. In the case of WSP, the optimization was carried out in three stages. In the first one, the parameters of Refs. \cite{perey,supersymmetry} were slightly readjusted to fit the 21.6 MeV energy angular distribution and then, used as a starting ansatz for the optimization procedure. The fits remarkably improved after the second stage, which consisted of optimizing the potential well depths. This quality enhancement was verified by the global estimator $\chi_{G}^{2}$ which changed from 13.1 to 1.57. In the last stage, the optimization included the optical potential geometrical parameters, which produced $\chi_{G}^{2}=1.18$. The optimization of the semi-microscopic SPP gave $\chi_{G}^{2}=1.97$, a very remarkable fact if one considers that it has only two degrees of freedom. 

Finally, the real part of the volume integral, calculated at an average sensitivity radius of 11.0 $\pm$ 0.5 fm, does not show the characteristic TA bump around the Coulomb barrier.  Our result differs from that of Ref. [36], where it was found that the real and imaginary parts of the potential follow the usual threshold anomaly behavior. This discrepancy indicates that more experimental data would be desirable in this region, as well as further theoretical analysis for deuteron scattering onto different targets.

As a future plan, we are making an exahustive analysis of the transfer and breakup reactions in the present system, where the elastic and inelastic breakup cross-section will be studied, as well as the role of the polarization potential.

\ack
The TANDAR accelerator crew is acknowledged. Brazilian authors acknowledge partial financial support from CNPq, FAPERJ, CAPES, and INCT-FNA (Instituto Nacional de Ci\^{e}ncia e Tecnologia - F\'{i}sica Nuclear e Aplica\c{c}\~{o}es) research project 464898/2014-5. The Argentinian authors acknowledge financial support from CONICET (Consejo Nacional de Investigaciones Cient\'ificas y T\'ecnicas) through Grant No. PIP0930, and from FONCyT (Fondo para la Investigaci\'on Cient\'ifica y Tecnol\'ogica) through Grants No. PICT-2017-4088 and PICT-2019-3565.

\providecommand{\noopsort}[1]{}\providecommand{\singleletter}[1]{#1}%

\end{document}